\documentclass{emulateapj}
\usepackage{psfig}

\newcommand{\be}{  \begin{eqnarray} }
\newcommand{\ee}{  \end{eqnarray}} 
\newcommand{\bd}{  \begin{displaymath} }
\newcommand{\ed}{  \end{displaymath} }
\newcommand{\msun}{ M_{\odot}}

\newcommand{\sigmat}{\sigma_{\rm T}}
\font\gkvec=cmmib10 
\def\bbeta{\hbox{{\gkvec\char12}}}  

\begin{document}
\title{The Effects of Magnetic Fields and Inhomogeneities on Accretion
Disk Spectra and Polarization}
\author{Shane W. Davis\altaffilmark{1,2}, Omer M. Blaes\altaffilmark{3},
Shigenobu Hirose\altaffilmark{4}, and Julian H. Krolik\altaffilmark{5}}

\altaffiltext{1}{School of Natural Sciences, Institute for Advanced Study,
Einstein Drive, Princeton, NJ 08540}
\altaffiltext{2}{Chandra Fellow}
\altaffiltext{3}{Department of Physics, University of California, Santa
Barbara, CA 93106}
\altaffiltext{4}{Institute for Research on Earth Evolution,
JAMSTEC, Yokohama, Kanagawa 236-0001, Japan}
\altaffiltext{5}{Department of Physics and Astronomy, Johns Hopkins University,
Baltimore, MD 21218}

\begin{abstract}
We present the results of one and three-dimensional radiative transfer
calculations of polarized spectra emerging from snapshots of radiation
magnetohydrodynamical simulations of the local vertical structure of
black hole accretion disks.  The simulations cover a wide range of
physical regimes relevant for the high/soft state of black hole X-ray
binaries.  We constrain the uncertainties in
theoretical spectral color correction factors due to the presence of
magnetic support of the disk surface layers and strong density
inhomogeneities.  For the radiation dominated simulation, magnetic
support increases the color correction factor by about ten percent, but this
is largely compensated by a ten percent softening due to inhomogeneities.
We also compute the effects of inhomogeneities and
Faraday rotation on the resulting polarization.  Magnetic fields
in the simulations are just strong enough to produce significant
Faraday depolarization near the spectral peak of the radiation field.
X-ray polarimetry may therefore be a valuable diagnostic of accretion
disk magnetic fields, being able to directly test simulations
of magnetorotational turbulence.
\end{abstract}


\section{Introduction}
\label{intro}

The most well understood accretion flow state onto a black hole is
that of a geometrically thin, optically thick disk.  Considerable
theoretical effort has been devoted to calculating spectral models of
such disks for comparison with observations (e.g.
\citealt{ks84,ln89,st95,sk98,hub01,dav05,hkh05,dh06}).  These models
have been compared to observed colors and spectra of active galactic
nuclei (e.g. \citealt{lao90,bon07,dwb07}), ultraluminous X-ray sources
\citep{hk08}, and black hole X-ray binaries \citep{ddb06,dd08}.  In
the last case, where the models appear to perform quite well for the
high/soft state, attempts have been made to use them to measure the
spins of black holes from observed X-ray continuum spectra
\citep{sha06,ddb06,mcc06,mid06,liu08,gou09}.

It is important to bear in mind, however, that all such models are
based on certain {\it ad hoc} assumptions.  First, they generally
adopt radial profiles of surface mass density and emissivity based on
some form of the alpha stress prescription of \citet{ss73} and some
inner boundary condition on the stress (usually zero).  These models also
make various assumptions about the vertical structure of the disk at
each radius.  In particular, at a given radius the structure depends
only on height above the midplane, and inhomogeneities generated by
instabilities and turbulence are neglected.  The disk is usually
assumed to be vertically supported against the tidal gravity of the
black hole by gas and radiation pressure alone.  Some prescription for
the vertical distribution of dissipation is adopted, e.g. that the
dissipation rate per unit mass is constant.  Finally, some mechanism
for vertical heat transport is assumed, e.g. radiative diffusion
and/or convection.

\begin{deluxetable*}{cccccccc}
\tabletypesize{\scriptsize}
\tablecaption{Parameters of Simulation Snapshots \label{tbl-1}}
\tablewidth{0pt}
\tablehead{
\colhead{Simulation} & \colhead{$M$} & \colhead{$r/r_{\rm G}$} &
\colhead{$\Omega$} & \colhead{Epoch} & \colhead{above or below} &
\colhead{$m_0$} & \colhead{$T_{\rm eff}$} \\
\colhead{} & \colhead{$(M_\odot)$} & \colhead{} &
\colhead{(rad s$^{-1}$)} & \colhead{(orbits)} & \colhead{midplane} &
\colhead{(g cm$^{-2}$)} & \colhead{(K)}\\
}
\startdata
0326c\tablenotemark{(a)} & 6.62 & 300 & 5.9 & 60 &
below & $5.10\times10^4$ & $5.25\times10^5$ \\
0528a\tablenotemark{(b)} & 6.62 & 150 & 17 & 90 &
above & $4.75\times10^4$ & $1.16\times10^6$ \\
1112a\tablenotemark{(c)} & 6.62 & 30 & 190 & 200 &
below & $5.37\times10^4$ & $3.45\times10^6$ \\
\enddata

\tablenotetext{a}{\citet{hks06}}
\tablenotetext{b}{\citet{khb07,bhk07}}
\tablenotetext{c}{\citet{hkb09}}

\end{deluxetable*}

Significant progress in understanding the vertical structure of
accretion disks has been made with vertically stratified shearing box
simulations of magnetorotational turbulence \citep{bh98}.  In
particular, radiation magnetohydrodynamical simulations of the
vertical structure have now been done for a broad range of midplane
radiation to gas pressure ratios of relevance to black hole accretion
disks \citep{tur04,hks06,khb07,bhk07,hkb09}.  These simulations have
demonstrated that magnetic forces contribute significantly to the
vertical hydrostatic balance of the disk, and can be dominant over gas
and radiation pressure gradient forces in the surface layers where the
spectrum is formed.  At the same time, these magnetically dominated
regions are Parker unstable, and as a result, very large density
inhomogeneities form near the thermalization and scattering
photospheres.  The mass density falls off steeply with height above
the midplane, and most of the accretion power is (numerically)
dissipated at high optical depth near the midplane regions.  Radiative
diffusion completely dominates the vertical heat transport, and the fraction
of accretion power carried out to the photospheres by Poynting or
mechanical fluxes is negligible.  Finally, the overall turbulent stress
scales approximately with the total thermal pressure near the
midplane.  Nonetheless, the inner radiation pressure dominated regions
of the disk are thermally stable \citep{tur04,hkb09}.  They may, however,
be subject to inflow instability \citep{le74}.

In previous work \citep{dav05,bla06}, we incorporated the numerical
dissipation profiles computed in the simulations of \citet{tur04} and
\citet{hks06} in one dimensional non-LTE vertical structure models of
disk annuli.  For models which neglect magnetic support, we found very
little difference in the emergent spectra compared to models based on
the assumption of constant dissipation per unit mass, the common
prescription which gives constant density in radiation dominated
regions.  Even though the numerical dissipation profiles have the
dissipation per unit mass generally increasing outward near the
photosphere, most of the accretion power is still dissipated and
thermalized at high optical depth, resulting in very similar spectra.

In contrast, the addition of substantial magnetic support of the outer
layers of the disk results in significant differences in the emergent
spectra \citep{bla06}.  Compared to models that neglect magnetic
forces \citep[e.g.][]{dav05}, magnetically supported atmospheres have
larger density scale heights, resulting in smaller densities at the
photosphere.  This increases non-LTE effects and the ionization state
of the gas, and also increases the ratio of scattering to absorption
opacity.  All of these effects result in a hardening of the disk
spectrum \citep{bla06,bp07}.

However, the strong density inhomogeneities that are seen in the
surface layers may act to soften the spectrum, both by increasing
the thermalization of the radiation through the enhanced absorption
opacity in the denser regions, and by enhancing the rate of cooling as
photons diffuse faster through the low density regions
\citep{dav04,beg06}.

The fact that accretion disks in black hole X-ray binaries in the
high/soft state are expected to be electron scattering dominated
implies that their thermal spectra should be significantly polarized
\citep{cha60}.  However, the complex photospheric geometry that
results from the strong density inhomogeneities seen in the
simulations may reduce the degree of polarization \citep{cs90}.  Even
if the disk surface is flat, relativistic effects can dilute the
polarization \citep{sc77}.  On the other hand, relativistic effects
can also cause photons emitted by one part of the disk to scatter off
a different part of the disk, enhancing the polarization and rotating
its angle \citep{ak00,sk09}.  Strong photospheric magnetic fields can
also Faraday depolarize the emergent radiation field \citep{gs78}, and
the fields seen in the simulations are marginally strong enough to do
just that \citep{bhk07}.  Future X-ray polarimetry satellites may
therefore provide valuable diagnostics of surface magnetic fields in
black hole accretion disks \citep{gss06,sil07}, providing a good
observational test of the simulation results.

In this paper we present the results of spectropolarimetric radiative
transfer calculations through representative simulation domains that
cover the range of radiation to gas pressure regimes that are relevant
to black hole X-ray binaries.  We quantify the degree of spectral
hardening due to magnetic support, and the degree of spectral
softening due to density inhomogeneities.  We also calculate the
expected polarization signatures of both of these effects, and confirm
that X-ray polarimetry will indeed be able to provide interesting
diagnostics on black hole accretion disk magnetic fields.

This paper is organized as follows.  In \S\ref{sims} we briefly
describe the numerical simulation data that we use in our radiative
transfer calculations.  In \S\ref{tlusty} we compute one-dimensional
non-LTE atmosphere models based on horizontally averaged dissipation
and magnetic force profiles taken from the simulations.  Then in \S
\ref{mc} we incorporate the fully three-dimensional structure of the
simulations by using Monte Carlo radiative transfer calculations to
compute the local emergent spectrum.  We keep track of the
polarization in these calculations, and we discuss the emergent
polarization spectrum in \S \ref{mcpol}.  Using certain scalings that
we derive from comparing the different simulations, we present an
illustrative full disk spectrum of the polarization in \S
\ref{diskpol}.  In \S \ref{discus} we discuss our results, in
particular quantifying the uncertainties in spectral color correction
factors and discussing how X-ray polarimetry could be used as a
diagnostic of disk magnetic fields.  We summarize our conclusions in
\S \ref{conc}.  Our Monte Carlo calculations fully incorporate
polarized Compton scattering, and we summarize the relevant equations
and methods in an Appendix.  Readers interested mainly in
observational applications of our work may wish to begin by reading \S
\ref{discus} and \S \ref{conc}.

\section{Simulation Data}
\label{sims}

We use data from three vertically stratified shearing box simulations
in this paper: a simulation of a gas pressure dominated box (0326c,
\citealt{hks06}), a box with comparable midplane gas and radiation
pressures (0528a, \citealt{khb07,bhk07}), and a radiation pressure
dominated simulation (1112a, \citealt{hkb09}).  The reader should
consult these papers for details of these simulations.  We summarize
the most relevant parameters for spectral modeling in Table 1.  In
particular, $M$ is the mass of the black hole, assumed to be
non-spinning; $r$ is the radius in the disk where the box would be
located, in units of the gravitational radius $r_{\rm G}\equiv
GM/c^2$; and $\Omega$ is the local angular velocity of the flow, used
to approximate the local tidal gravitational acceleration through
$|g|=\Omega^2|z|$, where $z$ is the height above the midplane.
Because we are interested in the full three dimensional structure of
the medium, we do not use time-averages but instead select three
representative epochs in each simulation.  Table 1 also shows the
column mass density $m_0$ to the midplane and the effective
temperature $T_{\rm eff}$ of the local emergent flux at that epoch as
computed by the simulation.

\begin{figure}
\begin{center}
\hbox{
\psfig{figure=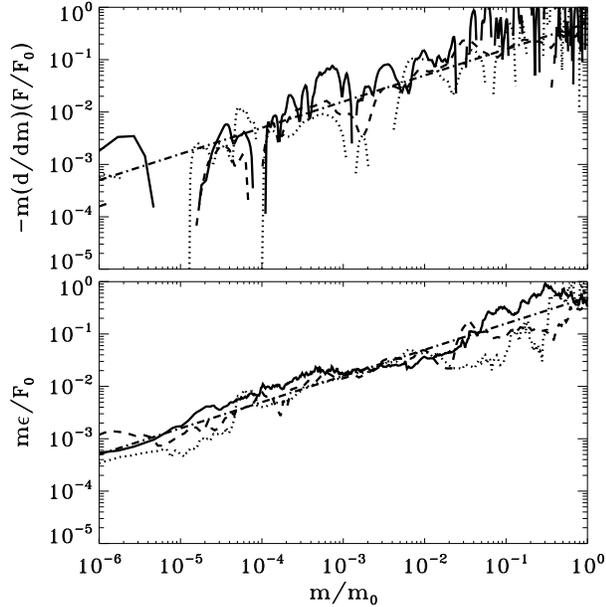,width=0.5\textwidth,angle=0}
}
\end{center}
\caption{Horizontally-averaged, vertical dissipation profiles for the
particular epochs of the three simulations we use in this paper.  In
the top panel we plot the column mass density derivative of the
vertical radiative flux, scaled with the total emergent flux
$F_0\equiv\sigma T_{\rm eff}^4$, as a function of column mass density $m$
scaled with the total column mass density $m_0$.  The dashed, dotted,
and solid curves correspond to simulations 0326c, 0528a, and 1112a,
respectively.  The bottom panel is the same as the top except that we
plot the normalized, horizontally averaged numerical dissipation per
unit mass $\epsilon$, as computed in the simulations. The dot-dashed
lines in each panel indicate a power law with exponent $1/2$, which
corresponds to the $dF/dm$ profile assumed in the TLUSTY calculations.
\label{f:mdfdm}}
\end{figure}

Figure \ref{f:mdfdm} shows the distribution of column mass times the
dissipation per unit mass as a function of column mass $m$ for each of
the three simulation data sets.  We measure this in two ways.  For the
upper panel we difference the vertical radiative flux with respect to
column mass, and for the lower panel we show the actual numerical
dissipation rate per unit mass $\epsilon$.  If local radiative
equilibrium held exactly, then these two quantities should have
identical distributions.  This is not exactly true in the simulations,
because a small fraction of the dissipated heat is transported
vertically outward through advection of internal energy \citep{hkb09}.
However, this is most important near the midplane, and is completely
negligible in the low column mass density regions where the emergent
spectrum is formed.

We have scaled the data shown in Figure~\ref{f:mdfdm} with the
midplane column mass density $m_0$ and the emergent radiative flux
$F_0$ for each simulation. With this normalization, it
is remarkable how similar all the distributions are, despite the
different physical conditions in each simulation and the different
numerical resolutions in the vertical direction.  Within the noise,
all the dissipation profiles are well fit by a power law of the form
\be
m\epsilon=-m\frac{dF}{dm}=\frac{F_0}{2}\left(\frac{m}{m_0}\right)^{1/2}.
\label{eq:dissprofile}
\ee

Note that this differs slightly from the broken power law profile used
in \citet{bla06} (Eq. 1 in that paper), which was based on a fit to
the dissipation profile computed from the 0326c simulation, but is
equivalent to the scaling derived by \citet{khb07} for the 0528a
simulation and used for the initial condition of the radiation
dominated run 1112a \citep{hkb09}.  \citet{bla06} found $mdF/dm\propto
m^{0.4}$ at high column mass densities and $\propto m$ at low column
mass densities.  Since this is only an approximate relation, and
subject to differences in averaging of the simulation profiles, the
small difference in the exponents is not significant.  We adopt the
1/2 exponent because it is the best match to the high column density
behavior in all three simulation, to within the uncertainties.  It is
important to recognize that all these dissipation profiles have most
of the dissipation occurring at high column mass densities, much
deeper than the spectral forming regions of the annuli.

\begin{figure}
\begin{center} \hbox{
\psfig{figure=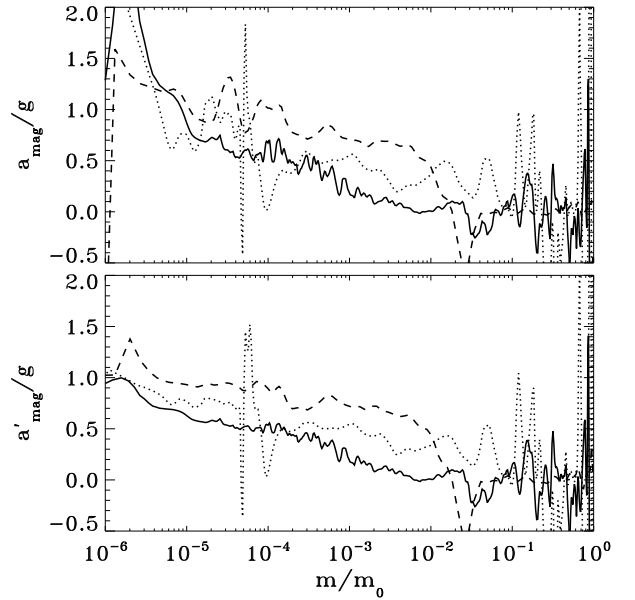,width=0.5\textwidth,angle=0} }
\end{center}
\caption{
Horizontally-averaged, vertical magnetic acceleration $a_{\rm
mag}$ as computed in the simulations (top) and renormalized magnetic
acceleration $a_{\rm mag}^\prime$ (bottom), scaled with the local tidal
gravitational acceleration $g$, as a function of column mass density $m$
scaled with the midplane column mass density $m_0$.  As in
Fig. \ref{f:mdfdm}, the dashed, dotted, and solid curves
correspond to simulations 0326c, 0528a, and 1112a, respectively. 
\label{f:magacc}}
\end{figure}

Figure \ref{f:magacc} shows the horizontally-averaged profiles of
magnetic acceleration $a_{\rm mag}$, scaled with the local
gravitational acceleration $g$, as a function of scaled column mass
density from the three simulation data sets.  The magnetic
acceleration is the net result of magnetic pressure and tension
forces, which are comparable in magnitude in the low column mass
density regions.  Near the surface, the magnetic acceleration can be
larger than the gravitational acceleration (even after averaging) due
to deviations from hydrostatic equilibrium, and even be negative (inward).

Such excursions may give rise to interesting variability, but we are
primarily interested in the time averaged behavior.  In an attempt to
better enforce hydrostatic equilibrium, we rescale the magnetic acceleration
via
\be
a'_{\rm mag}=\frac{a_{\rm mag}}{a_{\rm tot}}g,
\label{eq:amagprime}
\ee
where $a_{\rm tot}$ is the total acceleration due to radiation,
magnetic fields, and gas pressure gradients in the simulation.  This
rescaling is also shown in Figure \ref{f:magacc}.  It significantly
improves the agreement between the vertical density profiles of
the simulation data and those produced in our one dimensional atmosphere
models, to which we now turn.

\begin{figure}
\begin{center}
\hbox{
\psfig{figure=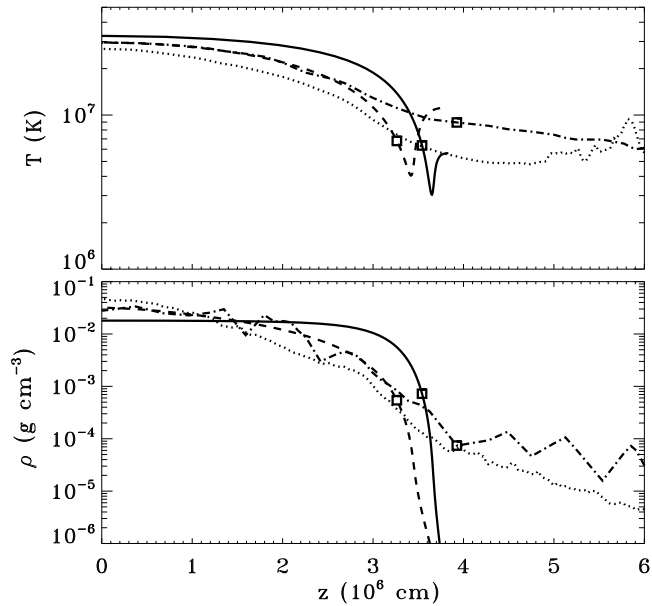,width=0.5\textwidth,angle=0}
}
\end{center}
\caption{
Temperature (top) and density (bottom) as a function of height above
the midplane for an annulus.  The input parameters have been chosen to
match the $t=200$ epoch of the 1112a simulation.  The solid curve
corresponds to a standard model, in which the dissipation rate per
unit volume is locally
proportional to density and magnetic support is neglected.  The
dashed curve also neglects magnetic pressure support, but
incorporates the simulation-based
dissipation profile of equation (\ref{eq:dissprofile}).  The dot-dashed
curve includes both the modified dissipation profile and the magnetic
support.  For each curve, the squares denote the location of the
effective photosphere
at energy 2 keV, which is near the peak in $\nu F_\nu$ for the spectra
shown in Fig. \ref{f:s1112}. For comparison, we also plot horizontally
averaged density and temperature profiles from the simulation (dotted
curves).
\label{f:mod1112}}
\end{figure}

\section{1d Spectral Models:  the effects of dissipation and magnetic support}
\label{tlusty}

We use the TLUSTY code \citep{hl95} to compute one dimensional
vertical structure models and emergent spectra from annuli whose
input parameters ($m_0$, $T_{\rm eff}$,
and $\Omega$) correspond to the three snapshots summarized in Table
\ref{tbl-1}.  We also incorporate the vertical dissipation and magnetic
support profiles measured from the three simulation epochs.

The effects of the vertical support and dissipation
profile are qualitatively similar in all three cases, so we only plot
the model corresponding to 1112a in Figure \ref{f:mod1112}.  (A
similar plot for the 0326c snapshot can be found in Fig. 6 of
\citealt{bla06}.)

The solid curve shows the ``standard'' annulus with constant dissipation
rate per unit mass (i.e. constant $dF/dm$)
and no magnetic support.  Since the annulus is radiation pressure
dominated throughout most of its extent, the density is very nearly
constant, as is commonly assumed.  However, the effective optical
depth of unity (marked by the squares) shows that the spectral forming
layer occurs close to the surface where gas pressure gradients
dominate and balance gravity to ensure hydrostatic equilibrium.  As
discussed in previous work \citep{dav05,dh06,ddb06,bla06,dd08}, this
means that the spectral properties are sensitive to the gas pressure
scale height near the surface, but rather insensitive to midplane
properties.

\begin{figure}[h]
\begin{center}
\hbox{
\psfig{figure=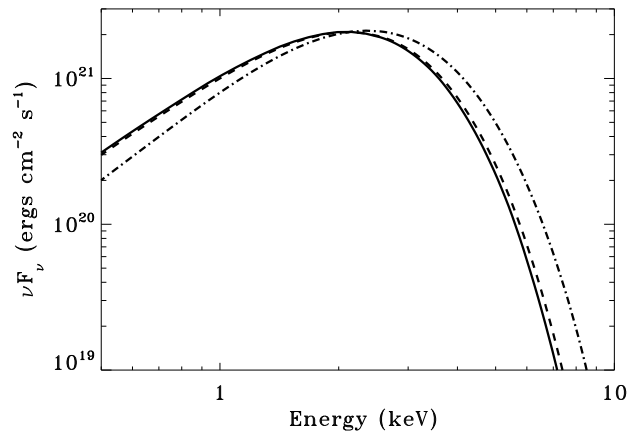,width=0.5\textwidth,angle=0}
}
\end{center}
\caption{Local emergent spectrum for an annulus viewed from an
inclination of $55^\circ$ to the surface normal. The annulus parameters
have been chosen to match the $t=200$ snapshot of the 1112a simulation 
($P_{\rm gas} < P_{\rm rad}$). The curves correspond to
emission from models with the standard dissipation and no magnetic
support (solid curve), modified dissipation but no magnetic support
(dashed curve), and modified dissipation and magnetic support included
(dot-dashed curve).
\label{f:s1112}}
\end{figure}

\begin{figure}[h]
\begin{center}
\hbox{
\psfig{figure=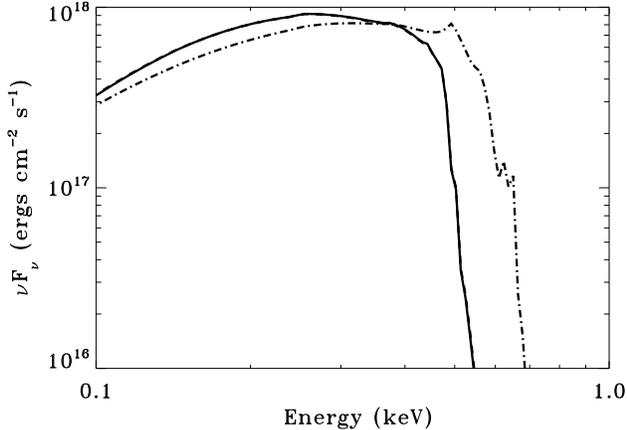,width=0.5\textwidth,angle=0}
}
\end{center}

\caption{Local emergent spectrum for an annulus viewed at an 
inclination of $55^\circ$ to the surface normal. The annulus parameters
have been chosen to match the $t=60$ snapshot of the 0326c simulation 
($P_{\rm gas} > P_{\rm rad}$).  The curves have the same
meanings as in Fig. \ref{f:s1112}. Note that the dashed and solid 
curves are nearly identical in this case.
\label{f:s060}}
\end{figure}

\begin{figure}[h]
\begin{center}
\hbox{
\psfig{figure=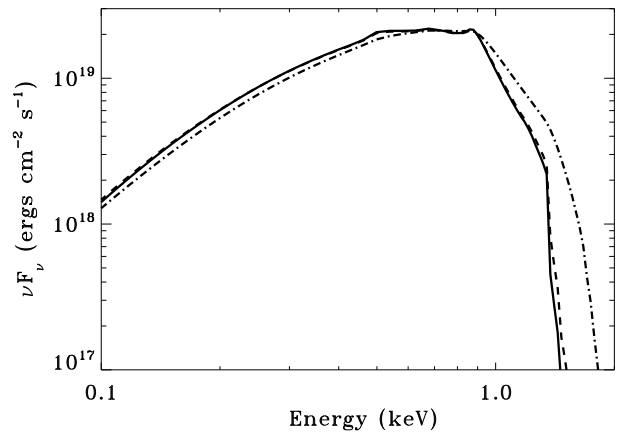,width=0.5\textwidth,angle=0}
}
\end{center}
\caption{Local emergent spectrum for an annulus viewed from an
inclination of $55^\circ$ to the surface normal. The annulus parameters
have been chosen to match the $t=90$ snapshot of the 0528a simulation 
($P_{\rm gas} \sim P_{\rm rad}$). The curves have the same
meanings as in Fig. \ref{f:s1112}.
\label{f:s090}}
\end{figure}

A comparison of Figures \ref{f:mod1112} and \ref{f:s1112} demonstrate
this lack of sensitivity.  The dashed curve corresponds to a model
that includes the modified dissipation profile, but still neglects
magnetic support.  Figure \ref{f:mod1112} shows a significant change
in the density profile, which is no longer constant at large depths
and yields a higher central density.  Also, the temperature is
generally lower at a given height beneath the effective photosphere
relative to the standard model.  However, the observed spectra in
Figure \ref{f:s1112} remain very similar with only a small amount of
hardening for the model with enhanced surface dissipation.  This can
be understood by noting that the temperature and density at the
effective photosphere (denoted by the squares) both occur near the
surface where the density declines rapidly.  Because the surface
pressure scale heights are still very similar for the two models, the
spectra are relatively unchanged even though the midplane properties
differ.  This is generically the case as long as only a small fraction
($\lesssim 10 \%$) of the dissipation occurs above the effective
photosphere \citep[see][ for further discussion]{dd08}.  All three of
the snapshot models obey this constraint, and inspection of Figures
\ref{f:s1112}, \ref{f:s060}, and \ref{f:s090} shows that modified
dissipation implied by the simulations has only a small effect on the
spectrum.

The model including magnetic support is shown as the
dot-dashed curve in Figure \ref{f:mod1112}.  The addition of magnetic
support greatly increases the pressure scale height near the surface.
The resulting profiles are in qualitative agreement with the
horizontally averaged profiles taken directly from the simulation
(dotted curves), although they are not as smooth as the simulation
results.  Also, the density is not strictly monotonic decreasing as
required for stable hydrostatic equilibrium.  These inconsistencies
result from the assumption of hydrostatic equilibrium in the
atmosphere calculation, even though the simulations can have
significant, non-hydrostatic dynamics near the surface.  Also, the
temperature differences arise partly due to differences in radiative
equilibrium in the two cases.  The radiation in the simulations is
handled through flux-limited diffusion \citep{lp81} and grey opacity
which only accounts for free-free emission and absorption.  The 1d
calculation solves full radiative transfer with bound-free opacities,
but neglects the effects of inhomogeneities in the full 3d geometry of
the simulations.  We'll expand on this point in \S\ref{mcspec}.

Due to the much larger pressure scale height at the surface, the
density (temperature) at the effective photosphere is much lower
(higher).  As discussed in \citet{bla06}, this combination of higher
temperature and lower density leads to harder spectra, both due to
the effects on absorption features at lower temperatures and due to
the effects of electron scattering at higher temperatures.  Indeed,
Figures \ref{f:s1112}, \ref{f:s060}, and \ref{f:s090} show that
magnetically supported annuli all give harder spectra (dot-dashed
curves), demonstrating that the qualitative result (harder spectra)
is quite general.  We can make this more quantitative by comparing the
color corrected blackbody models that best match the magnetized and
standard atmospheres.  This estimate is most robust for the model
corresponding to the 1112a snapshot since the strong absorption edges
in the other models prevent the color-corrected blackbody from
providing a good fit, making the color correction (or hardening
factor) somewhat ambiguous.  For the 1112a model, which lacks
significant edge features, we find the color correction increases from
about 1.67 to 1.84 when incorporating magnetic support, a roughly
10\% increase.

\section{3d Spectral Models: the effects of inhomogeneities and
magnetic fields}
\label{mc}

\subsection{Monte Carlo Radiative Transfer}

Although the hydrostatic, 1d calculations described in \S\ref{tlusty}
capture much of the most relevant physics, they neglect all the 3d,
dynamical information available from the simulation results.  In
order to explore these effects we have computed fully 3d radiative
transfer calculations using Monte Carlo (MC) methods for each snapshot
listed in Table \ref{tbl-1}.

The MC code reads in a grid of densities, temperatures, and magnetic
field vectors from a single 3d snapshot of a simulation.  These
variables are taken to be constant in time throughout the MC
calculation.  At the beginning of the calculation, the electron
density, free-free emissivity and absorptivity are computed for each
zone.  We assume the plasma is completely ionized with a ten percent
number abundance of helium.  (The MC calculations do not account
for partial ionization, non-LTE effects or bound-free processes.
These {\it are} included in the 1d calculations of \S\ref{tlusty}.)

\begin{figure}[h]
\begin{center}
\hbox{
\psfig{figure=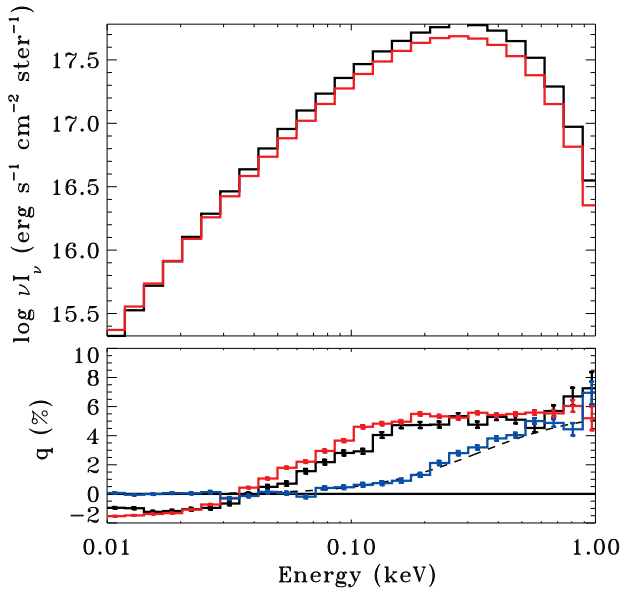,width=0.5\textwidth,angle=0}
}
\end{center}

\caption{
Local emergent intensity (top) and polarization (bottom) from the
$t=60$ snapshot of the 0326c simulation ($P_{\rm gas} > P_{\rm rad}$),
viewed at an inclination of $79^\circ$ to the surface normal.  In the top
panel, the spectra are generated from fully 3d (black curve) and
horizontally averaged 1d (red curve) Monte Carlo calculations. In the
bottom panel, we compare the polarization from 1d calculations (red
curve), and 3d calculations with (blue curve) and without (black
curve) the effects of Faraday rotation.  The dotted curve in the bottom
panel corresponds to fitting function given by eq. \ref{eq:qfit}.
\label{f:mc060}}
\end{figure}

The code then propagates photon packets through the domain until they
escape or are absorbed.  The domain is assumed to be periodic in
both the radial and azimuthal directions\footnote{The simulations
themselves assume shearing periodic boundary conditions in the radial
direction \citep{hgb95}.  Because we neglect time evolution,
we simply assume periodic boundary conditions in the radial direction
for the Monte Carlo calculation.},
so escape only occurs through the vertical boundary.  For
efficiency, we only include the outer portions of the full disk volume near
the surface, and assume reflecting boundaries at some inner base.
We choose this inner boundary to be at a large optical depth to true absorption
($\gtrsim 10$ for typical energies) so that its location
has negligible effect on the output spectra and polarization.

The initial location of each packet is chosen randomly throughout the
domain, and each packet is assigned a weight based on the emissivity
of the initial grid zone.  The initial direction vector of the packet
is randomly chosen from an isotropic distribution.  The photons are
initially unpolarized and the energy is randomly chosen and assigned a
weight to approximate the frequency dependence of the free-free
emissivity.

After initialization, the absorption and electron scattering mean free
paths are calculated for the appropriate energy and the packet is
advanced along a ray pointing along the direction vector by an optical
depth chosen randomly from an exponential distribution.  If this
optical depth is sufficiently large, the packet may be propagated into
a neighboring grid zone.  The code assumes each grid zone is
homogeneous in density, temperature, and magnetic field strength. The
mean free paths are updated each time the packet enters a new grid
zone, and propagation continues until the packet moves the prescribed
optical depth or exits the domain.  If the packet escapes, its
weighted contribution is added to the output arrays corresponding to
its final direction and energy, and a new packet is initialized.

If the packet remains in the domain, its weight is reduced by the
ratio of absorption to total (absorption and electron scattering)
opacity. If the packet weight falls below a prescribed minimum weight,
it is considered absorbed and a new photon packet is initialized and
propagated.  (The minimum weight is chosen to be sufficiently low that
the absorbed photons have negligible effect on the output spectrum.)
Otherwise, the packet is assumed to have scattered and the energy,
direction vector, and polarization vector are updated.  We model the
scattering process using the MC methods for Compton scattering as
described in \cite{pss83}, but with modifications to include
polarization as described in the appendix.  A new optical depth is
then drawn and the process is repeated until the packet escapes or is
absorbed.

\begin{figure}[h]
\begin{center}
\hbox{
\psfig{figure=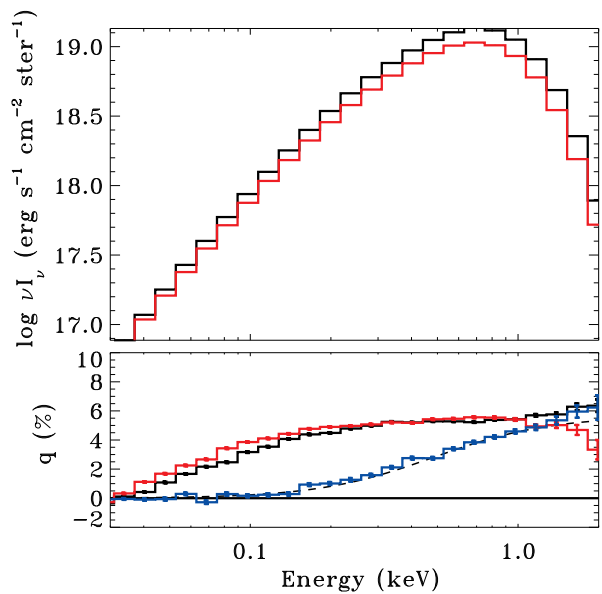,width=0.5\textwidth,angle=0}
}
\end{center}
\caption{
Local emergent intensity (top) and polarization (bottom) from the
$t=90$ snapshot of the 0528a simulation ($P_{\rm gas} \sim P_{\rm
  rad}$), viewed at an inclination of $79^\circ$ to the surface normal.
The dotted and binned curves have the same meaning as in Fig.
\ref{f:mc060}.
\label{f:mc090}}
\end{figure}

In order to test our MC routines we ran a number of comparison
calculations with other methods.  Since we did not have a single
comparison method that could suitably handle all the complexity of the
MC calculation, we separated our test into two parts.  We first tested
the initialization and photon propagation routines by comparing with
1d Feautrier calculations that model polarized Thomson scattering and
allow for temperature and density variations in a single dimension.
We then ran our MC code with polarized Thomson scattering on fully 3d
grids with equivalent density and temperature variations in the
vertical dimension, finding good agreement for both the specific
intensity and polarization spectra.  We next
tested the polarized Compton scattering routines by comparing with
results from a code \citep{hb98} based on the iterative scattering
method of \citet{ps96}.  We used both methods to calculate the
emergent specific intensity and polarization from a homogeneous,
finite optical depth ``corona'' with a blackbody seed photon
distribution at its base, finding good agreement.

\begin{figure}[h]
\begin{center}
\hbox{
\psfig{figure=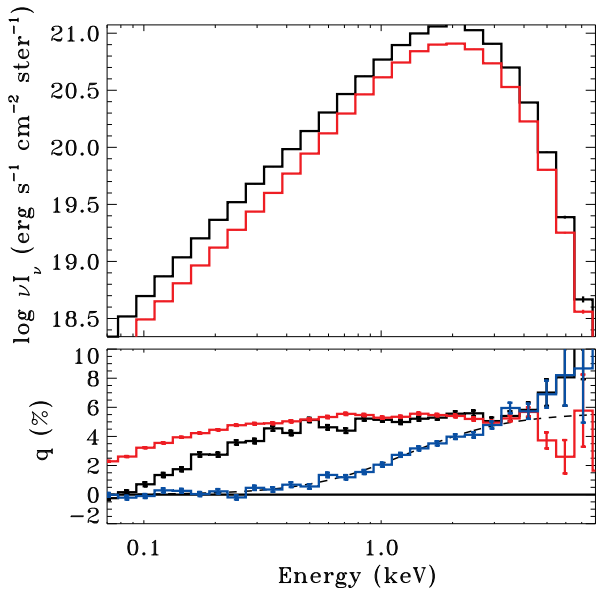,width=0.5\textwidth,angle=0}
}
\end{center}

\caption{Local emergent intensity (top) and polarization (bottom)
from the $t=200$ snapshot of the 1112a simulation ($P_{\rm gas} < P_{\rm rad}$), 
viewed at an inclination of $79^\circ$ to the surface normal.  The curves
have the same meaning as in Fig. \ref{f:mc060}.
\label{f:mc1112}}
\end{figure}

\subsection{The specific intensity:  effects of inhomogeneities}
\label{mcspec}

Figures \ref{f:mc060}, \ref{f:mc090}, and \ref{f:mc1112} show the
results of the MC calculations described above for the 0326c, 0528a,
and 1112a snapshots, respectively.  For the moment, we concentrate on
the top panels of each figure, which show the specific intensity at an
inclination of $79^\circ$.  As one would expect on physical grounds,
the specific intensity shows little azimuthal variation, so we plot
the azimuthally-averaged output of the MC calculation to improve the
statistics.  The solid black curves correspond to the results of MC
calculations through the full 3d grid and the red curves are the MC
spectra from horizontally averaged 1d grids.  In each case both sets
of calculations yield rather similar spectral shapes, and the spectral
peaks occur at nearly identical energies.  The main difference is that
the 3d model intensity is greater at all but the lowest energies.
Although we have only shown the spectra at $79^\circ$, this result
qualitatively holds at all inclinations, implying that a larger flux
is being radiated in the 3d calculations, in good agreement with
previous results \citep{dav04}.

This flux enhancement is primarily the result of the density
inhomogeneities.  Although there are some horizontal variations in the
temperature, they are generally smaller than the variations in the
density.  The weaker variation in temperature results from the
relatively short photon diffusion time scale, which efficiently
smooths out temperature fluctuations \citep[see e.g.][]{tur05}.  The
large density inhomogeneities are a result of highly variable magnetic
forces that dominate the surface layers.  Densities range over factors
of roughly ten above and below the mean at the horizontally averaged
effective photospheres in all the simulations \citep{bhk07,hkb09}.  It
is possible that photon bubbles, if they exist, may make the density
even more inhomogeneous \citep{tur05} in the radiation pressure
dominated regime.

The density inhomogeneities enhance the total emission because
free-free emission (and thermal emission processes in general) scale
as the square of the density.  Therefore, even though average density
is identical in the two calculations, the excess emission from high
density regions more than compensates for the deficit in low density
regions.  Of course, free-free absorption also scales as the square of
the density and if it were the only opacity source, increased
absorption would offset the increased emissivity.  However, since
electron scattering is proportional to one power of density and
dominates the opacity at most frequencies, photons emitted in high
density regions scatter out to lower densities before they are
absorbed, increasing their likelihood of escape.

Ultimately, we would like to address whether or not these
inhomogeneities are essential to obtaining the correct specific
intensity.  The fact that both the 1d averaged domain and 3d domain
yield the same spectral shape suggests that the spectra may be
insensitive to the inhomogeneities.  However, the flux enhancements
suggest we might be able to get an equivalent flux with a lower
surface temperature, possibly lowering the spectral hardening.  The
problem with comparing the 1d horizontally averaged and full 3d MC
calculations is that they don't represent the same
radiative equilibrium. In a real disk the flux of radiation (and
therefore the radiative equilibrium) is fixed by the overall amount
and location of turbulent energy dissipation.  One needs to compare
the 3d calculation against a 1d calculation with the same radiative
equilibrium.  Fortunately, we have already described such calculations
in \S\ref{tlusty}.

\begin{figure}[h]
\begin{center}
\hbox{
\psfig{figure=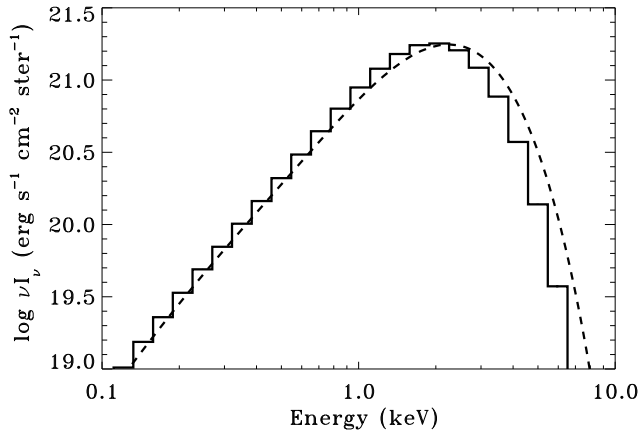,width=0.5\textwidth,angle=0}
}
\end{center}

\caption{Local emergent intensity at $55^\circ$
from the $t=200$ snapshot of the 1112a simulation ($P_{\rm gas} < P_{\rm rad}$), 
viewed at an inclination of $79^\circ$ to the surface normal.  The binned
curve is identical to the black binned curve in Fig. \ref{f:mc1112}. The
solid, unbinned curve is a TLUSTY calculation similar to the dot-dashed
curve in Fig. \ref{f:s1112}, but without bound-free opacity and with a
lower effective temperature as discussed in \S\ref{mcspec}.
\label{f:comp}}
\end{figure}

Figure \ref{f:comp} shows a comparison of the 3d MC spectrum (bins)
from Figure \ref{f:mc1112} with an equivalent TLUSTY atmosphere
spectrum (solid curve).  The TLUSTY model includes the simulation
derived dissipation profile and magnetic support, but differs from the
one shown in Figure \ref{f:s1112} (the dot-dashed curve) in that we
have dropped the bound-free opacity and used a slightly lower
effective temperature to better match the 3d MC calculation.  The
lower effective temperature is needed because the total flux in the MC
calculation is slightly lower than that found directly by the
simulation.  We find that the 3d MC spectrum is significantly softer than
the 1d TLUSTY spectrum.  In fact, the peak is lower by about 12\%,
almost completely offsetting the hardening due to the magnetic
support discussed in \S\ref{tlusty}!

In fact, a similar cancellation also occurs in the 0326c and 0528a 
snapshots as well.  Again, we repeated the 1d TLUSTY calculations 
from \S\ref{tlusty}, but neglecting bound-free opacities and fixing
the effective temperature to match the MC results.  The
spectral hardening due to magnetic support 
was about the same level as the softening due to inhomogeneities
in both cases: $\sim 10$\% for 0326c and $\lesssim 5$\% for 0528a.
Since the magnetic fields providing the support are also the
primary source of the inhomogeneities, a correlation between
the effects is not surprising.  However, it's remarkable
that the magnitude of the effects are such that they cancel out
in all three sets of calculations.

\begin{figure}[h]
\begin{center}
\hbox{
\psfig{figure=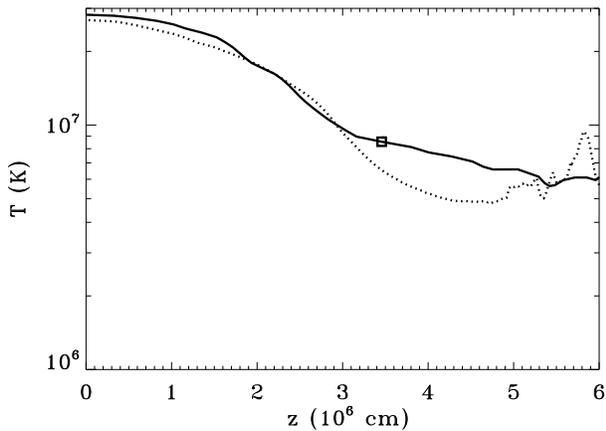,width=0.5\textwidth,angle=0}
}
\end{center}
\caption{
Temperature as a function of height above
the midplane for an annulus.  The solid curve is a 1d atmosphere
calculation with input parameters chosen to
match the $t=200$ epoch of the 1112a simulation, but with a slightly
lower effective temperature and neglecting bound-free opacity
in order to match the 3d MC calculation.  
For comparison, we also plot the horizontally
averaged temperature profile from the simulation (dotted
curve).
\label{f:temp}}
\end{figure}

We plot a comparison of the temperature profiles from the two
calculations in Figure \ref{f:temp}.  The dotted curve shows the
horizontally averaged temperature from the 1112a simulation snapshot
and the solid curve is the 1d atmosphere model.  Overall, the 1d
profile is a better match to the simulation than the equivalent model
in Figure \ref{f:mod1112} (dot-dashed curve) since we have modified
the input parameters to better match the MC calculation.
Nevertheless, the temperature gradients differ near the surface and
the effective photosphere (square symbol) is $\sim 15\%$ higher than
the simulation at the same height above the midplane.  We generated
similar plots for the other two snapshots and they are qualitatively
consistent with Figure \ref{f:temp}.

This result, along with the flux enhancements discussed above, strongly
suggests that the inhomogeneities allow the same flux to be radiated
with average lower surface temperatures, leading to a softer spectrum.
However, one must note that a number of other differences between the
two calculations may also contribute to the resulting discrepancies in
the temperature profiles and spectra.  The treatment of radiation transport
in the simulations (grey opacity, flux limited diffusion) differs from
both the MC and the TLUSTY
calculations.  Also, non-hydrostatic motions are significant in the
surface layers of the simulations while the 1d TLUSTY calculation
strictly enforces hydrostatic equilibrium.  Distinguishing between
these effects is difficult, but it is clear that the 3d and dynamic
nature of the simulations plays a non-negligible role in determining
the disk spectrum.  Ignoring these effects will lead to an
underestimate of the spectral hardening.

\section{Polarization:  the effects of Compton scattering, inhomogeneities,
and magnetic fields}
\label{mcpol}

For viewing angles inclined to the surface normal, homogeneous
atmospheres will appear moderately linearly polarized (up to 12\%) at
photon energies for which electron scattering is nearly elastic and
dominates the opacity \citep[see e.g.][]{cha60}.  Since electron
scattering opacity is typically dominant for the energy ranges of
interest, the MC calculations do produce significant linear
polarization at moderate to high inclinations.  This can be seen in
Figures \ref{f:mc060} - \ref{f:mc1112}, which plot the polarization
viewed from an inclination of $79^\circ$.

These polarization results are most easily discussed in terms of the
Stokes parameters corresponding to the total specific intensity $I$
and the linearly polarized intensities $Q$ and $U$.  (Electron
scattering does not impart circular polarization so the Stokes
parameter $V$ is identically zero in our calculations.)  We will also
find it convenient to define the total polarization $P \equiv
(Q^2+U^2)^{1/2}$.  In this paper we always plot the normalized
Stokes parameters ($q \equiv Q/I$, $u\equiv U/I$) and degree of
polarization $p\equiv(q^2+u^2)^{1/2}$.

We define $Q$ such that positive and negative $Q$ correspond
(respectively) to polarization perpendicular and parallel to the
surface normal.  $U$ corresponds to polarization defined relative to
axes rotated by $45^\circ$. For a homogeneous domain, $U$ would also
be identically zero by symmetry.  Although the simulation domains are
not homogeneous, they are still highly symmetric, and $U$ is nearly
zero in all three snapshots for all viewing angles.  Therefore, we
only plot $Q$ in Figures \ref{f:mc060} - \ref{f:mc1112}.  Like the
intensity, we do not observe significant azimuthal variation in the
polarization, so we plot only the azimuthal average.  The only
exception is when the effects of Faraday rotation are included, as
discussed in \S\ref{farad}.

The direction of the polarization is generally determined by the
angular dependence of the emergent radiation field.  Limb darkened
radiation yields polarization in the plane of the disk (positive $Q$),
isotropic radiation is unpolarized, and limb-brightened radiation
provides polarization parallel to the surface normal (negative $Q$).
Since Thomson scattering dominated atmospheres are limb darkened, one
nominally expects polarization with positive $Q$ \citep{cha60}.
However, a number of other processes are relevant: absorption at low
energies, changes in photon energy due to inelastic electron
scattering, Faraday rotation by the magnetic fields, and geometric
effects due to the inhomogeneities.

Absorption opacity changes the polarization both by competing with
scattering opacity and by altering the anisotropy of the radiation
field.  Depending on the vertical gradient of the thermal source
function, this can increase the polarization compared to a pure
scattering atmosphere \citep{har69,los79,bks85}, or decrease it and
even turn it negative \citep[the ``Nagirner effect'', c.f.][]{gs78}.
For the free-free opacity assumed here, absorption dominates at lower
photon energies, and the Nagirner effect is clearly evident in the
bottom panel of Figure \ref{f:mc060} at energies below 0.03~keV.
It is also present in the other two snapshots, but at lower energies
than those plotted in Figures \ref{f:mc090} and \ref{f:mc1112}.

\begin{figure}[h]
\begin{center}
\hbox{
\psfig{figure=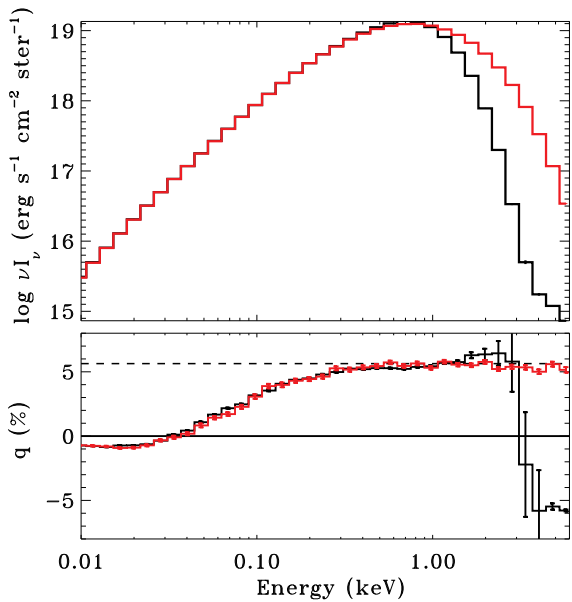,width=0.5\textwidth,angle=0}
}
\end{center}

\caption{Local emergent intensity (top) and polarization (bottom)
from the $t=90$ snapshot of the 0528a simulation ($P_{\rm gas} \sim P_{\rm rad}$), 
viewed at an inclination of
$79^\circ$ to the surface normal.  The black and red curves correspond
to full 3d MC calculations where electron scattering is treated
in the Compton and Thomson limits, respectively.  The dashed curve
shows the prediction for a semi-infinite, scattering-dominated
atmosphere viewed from the same inclination.
\label{f:thom}}
\end{figure}

\subsection{Compton scattering}

The effects of inelastic (Compton) electron scattering manifest
themselves as a turnover in $Q$ at photon energies well above the
spectral peak.  If electron scattering is treated in the elastic
(Thomson) limit the polarization would remain positive.  We have
deliberately neglected to plot this turnover in Figures \ref{f:mc060}
-\ref{f:mc1112} due to the poor statistics at these energies.  But,
the effect can be seen in Figure \ref{f:thom} where we compare a full
3d model where electron scattering is treated in the Compton
limit (black curve) with one in which the it's computed in the
Thomson limit (red curve).

At high energies above or near the peak, the predictions for
polarization differ significantly. The result is {\it not} due to the
differences in the scattering matrices, since the full
Compton-scattering matrix is well approximated by the Rayleigh matrix at
the energies of interest.  (See the appendix for further details.)
Instead, it is related to the different frequency dependence of the
angular distribution of the radiation in the two cases.  Since the
photon energies are fixed in the Thomson model, high energy photons
tend to be preferentially emitted deeper in the atmosphere where
temperatures are greater.  Since photons are initially emitted at
large scattering depths, the radiation field becomes limb darkened,
the spectrum becomes a modified blackbody \citep{rl79}, and the
polarization is well approximated by the predictions for a
homogeneous, semi-infinite, scattering-dominated atmosphere
\citep[see e.g.][]{cha60}.

However, when Compton effects are included these high energy photons
emitted at large depth can now efficiently exchange energy with the
electrons near the surface.  Since the surface is cooler, these
electrons are typically lower in energy and photons will tend (on
average) to lose energy until they have a mean energy similar to the
photospheric electrons.  This leads to the Wien tail at high photon
energies seen in the specific intensity for the Compton spectra.  The
photon angular distribution and the energy distribution are now
coupled by the scattering history, and the polarization can therefore
differ from that of an optically thick Thomson scattering atmosphere.
Indeed, the polarization of 2~keV photons is slightly enhanced by
Compton scattering in Figure \ref{f:thom}.

Near the surface photon diffusion smooths out most, but not all,
temperature inhomogeneities.  In a few localized regions the gas
temperature significantly exceeds the effective temperature of the
radiation, though this is likely artificial as simulation 0528a
neglected Compton energy exchange between the radiation and the gas
\citep{bhk07}.  In these pockets, photons are up-scattered to higher
energies, producing the break seen at high energy in the top panel of
Figure \ref{f:thom}.  Due to the small optical depth of the hot
material, these photons only overcome the emission in the Wien tail at
high energies.  At these energies the up-scattered emission is modestly
limb-brightened, due to the slightly increased probability of
laterally propagating photons to be scattered before escape.  This is
completely analogous to the polarization flip produced by inverse
Compton scattering in an optically thin, hot corona in plane parallel
geometry \citep{haa93,hb98}, and is a reminder that such coronal
geometries, if they exist in nature, will produce high energy
radiation that is polarized parallel to the projected disk rotation
axis.

\subsection{Inhomogeneities}

The effects of inhomogeneities can be seen in the bottom panels of
Figures \ref{f:mc060} - \ref{f:mc1112}.  The black curves show the
polarization of the full 3d calculations (excluding the effects of
Faraday rotation) while the red curves correspond to the 1d,
horizontally averaged domain.  We find that the 3d polarization curve
is generally less polarized than the 1d horizontally-averaged curve,
although the effect is rather small.

\begin{figure}[h]
\begin{center}
\hbox{
\psfig{figure=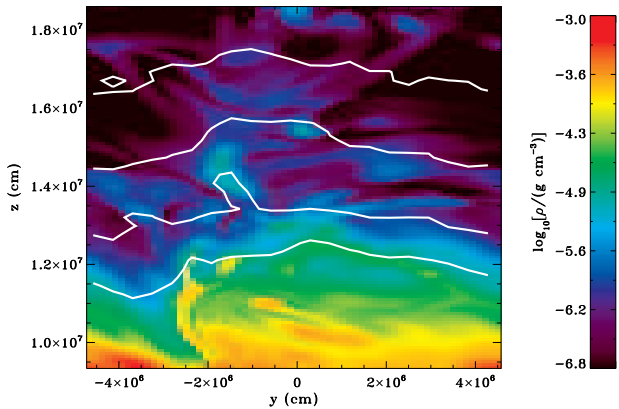,width=0.47\textwidth,angle=0}
}
\end{center}

\caption{Contours of constant escape fraction overplotted
on a two dimensional density slice in the $x-z$ plane of the
0528a snapshot.  The white contours correspond to
the fraction of upward moving photons that escape from a grid zone after
any scattering or emission of a packet.  From top to bottom,
they correspond to escape fractions $f=\exp(-\tau)$ with $\tau=0.5$, 1, 2, 
and 4.  
The plotted region corresponds to the uppermost 31\% of the
simulation domain.
\label{f:escf}}
\end{figure}

This slight reduction in polarization is in qualitative agreement with
the expectations of \citet{cs90}, who argued that deviations from a
planar photosphere (e.g. a corrugated surface) might produce a more
isotropic, or even limb brightened, radiation field.  They postulated
that such a photospheric geometry could, in principle, explain the
apparent results that AGN are at most weakly polarized ($\lesssim
2\%$) and that the polarization is perpendicular to the inferred disk
plane, contradicting the standard expectations of scattering dominated
atmospheres. However, the effect observed in our calculations is much
smaller than suggested by \citet{cs90} and may partly be the result of
the increased ratio of absorption to scattering opacity at some
energies.

The model of \citet{cs90} implicitly assumes the photosphere geometry
is complex, but that characteristic length scales of the horizontal
variations is significantly greater than the photon mean free path.
(If the horizontal length scale is smaller or comparable to a mean
free path, their assumption of a \citet{cha60} limb-darkening profile
relative to the local surface normal would be invalid.)  

To compare with their model we examined the shape of the photosphere
in our 3d calculations.  In 1d atmospheres the photosphere corresponds
to a surface of constant escape fraction, a concept which is easily
generalized to 3d.  Therefore, we computed in each grid zone the
fraction of photon packets escaping after each scattering or emission
event.  In Figure \ref{f:escf} we plot the contours of constant escape
fraction as white curves on top of the density in a 2d slice of the
0528a snapshot.  Although deviations from a planar surface are
significant, the length of horizontal variations are comparable to or
smaller than a typical mean free path to electron scattering.
Therefore, the geometric effect discussed by \citet{cs90} is nearly
negligible for these simulation domains.  

Nevertheless, these results do not invalidate the \citet{cs90}
hypothesis for real systems.  Since periodic boundary conditions are
assumed, larger length scale horizontal variations are not possible in
the current simulations.  The surface inhomogeneities are primarily
the result of Parker instability \citep{bhk07}, so larger wavelength
variations may be possible in actual accretion flows, and this
hypothesis should be reexamined when larger domains become
computationally feasible.

\subsection{Magnetic fields}
\label{farad}

The bottom panels of Figures \ref{f:mc060} - \ref{f:mc1112} also show
the effects of magnetic fields on the polarization.  Both the blue and
black curves are 3d calculations, but with and without (respectively)
the effects of Faraday rotation included.  Each time a photon packet
is propagated through a Thomson optical depth $\tau_{\rm T}$, the
polarization vector is rotated by an amount
\be 
\chi_{\rm F} = \frac{3 \lambda^2 \tau_{\rm T}}{16\pi^2 e} \bf{B
  \cdot {\hat k}}
 \label{eq:farad} 
\ee 
where $\bf{B}$ is the magnetic field in the current grid zone,
$\bf{\hat k}$ is a unit vector in the direction of photon propagation, $e$
is the electron charge, and $\lambda$ is the wavelength. For the
observed polarization, the rotation that occurs after the last
scattering is typically most important, so $\tau \sim 1$.  Since
$\bf{\hat k}$ is a unit vector, the typical rotation angle only depends on
$\lambda$, $\bf{B}$, and the direction.  Due to the $\lambda^2$
dependence in equation (\ref{eq:farad}), one is normally not concerned
with Faraday rotation in X-ray sources.  However, the near
equipartition strengths of magnetic fields in the simulation ($\sim
10^5-10^6$ G) are large enough that Faraday rotation can be significant
even at these short wavelengths.

For any single photon, Faraday rotation can only change the direction
of the polarization, not its magnitude.  However, due to
departures from uniformity in the magnetic field and differences in
trajectories, different photons will experience rotation through
different angles.  This is particularly significant for low energies
where $\chi_{\rm F} \gg 1$ and small differences along nearby
trajectories can lead to rather different polarization angles.  As a
result, photons are imparted with a nearly uniform distribution in
polarization, yielding zero net polarization after integration.
Therefore, Faraday rotation tends to completely depolarize at low
energies, gives modest reductions at intermediate energies where
$\chi_{\rm F} \gtrsim 1$, and has negligible effect for higher
energies where $\chi_{\rm F} \ll 1$. 

\begin{figure}[h]
\begin{center}
\hbox{
\psfig{figure=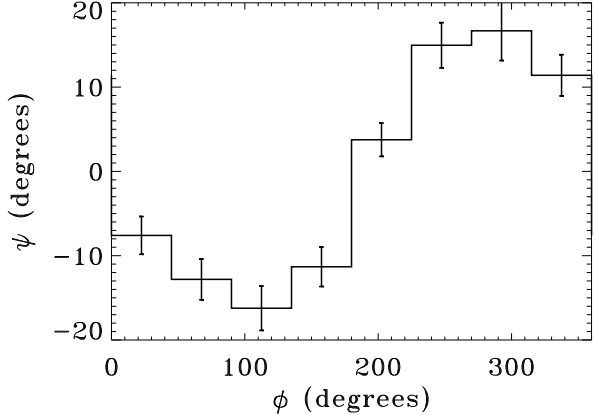,width=0.5\textwidth,angle=0}
}
\end{center}

\caption{Polarization angle as a function of azimuthal angle for the
$t=90$ snapshot of the 0528a simulation.  The polarization angle
is computed for an inclination of $79^\circ$, and averaged over photon
energies from 0.3-1 keV, where the Faraday rotation angle is near
unity.
\label{f:phi}}
\end{figure}

For photon energies with $\chi_{\rm F} \sim 1$, there is an
azimuthally dependent rotation, due to the nearly toroidal net
magnetic field in the snapshot.  Figure \ref{f:phi} shows the
polarization angle
\be
\psi\equiv0.5\tan^{-1}(U/Q)
\label{e:psi}
\ee
plotted as a function of azimuthal angle $\phi$ for the 0528a
snapshot.  A $\psi$ of zero corresponds to polarization parallel to
the simulation domain surface.  This plot shows the polarization angle
for an inclination of $79^\circ$, averaged over photon energies from
0.3-1 keV, where $\chi_{\rm F} \sim 1$.  For $\phi \sim 110^\circ$ and
$290^\circ$ the escaping photons are traveling parallel and
anti-parallel to the net toroidal field, obtaining non-zero
polarization angles.  When averaged over all azimuth, $\psi$ is
consistent with zero.

The observed spectrum from an unresolved disk is a weighted average
over all azimuth.  It is ``weighted'' because relativistic beaming can
enhance emission from the approaching side and diminish emission from
the receding side of a disk annulus.  Therefore, cancellation will not
be complete deep in the relativistic potential well of a black hole,
so some net rotation may be observed.  Nevertheless, we have chosen to
plot the uniformly azimuthally averaged $q$ in Figures \ref{f:mc060} -
\ref{f:mc1112}, since the details of the relativistic beaming are
complicated to model, and depend precisely upon where the simulation
would be located in a real accretion flow.  For higher or lower photon
energies, there is very little net rotation, and $U$ remains
consistent with zero, even before azimuthally averaging the output.

In the bottom panels of Figures \ref{f:mc060} - \ref{f:mc1112} 
the dashed curve corresponds to the simple analytic relation
\be
q=q_0(\mu)\left[1+(\lambda/\lambda_B)^2\right]^{-1},
\label{eq:qfit}
\ee
where $q_0(\mu)$ is the non-magnetized polarization value appropriate
for the observed inclination in a scattering dominated atmosphere
\citep{cha60}, $\lambda_B^2 \equiv 8\pi^3 e/(3 B_{\rm ph})$, $B_{\rm
  ph}$ is the average magnitude of $\bf B$ at the photosphere of the
snapshot, and $\mu$ is the cosine of the inclination angle.  The
quantity $(\lambda/\lambda_B)^2$ is roughly the Faraday rotation angle
for unit optical depth.  For energies near or below the peak, this
functional form provides a rather good approximation to the 3d MC
results when Faraday rotation is included.  It breaks down at high
energies because it does not account for the Compton scattering
turnover, but is a nearly perfect match to the Thomson scattering
calculations.

\begin{figure}[h]
\begin{center}
\hbox{
\psfig{figure=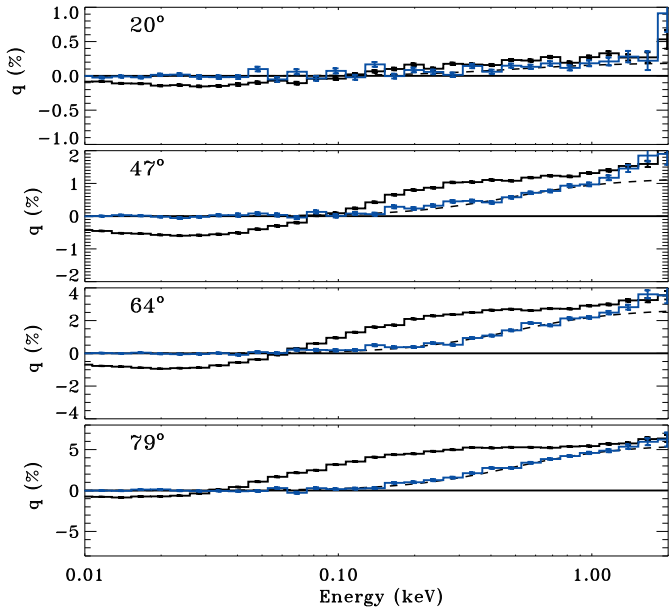,width=0.5\textwidth,angle=0}
}
\end{center}

\caption{Polarization viewed from four different inclinations
  to the surface normal. The spectra are generated from fully
  3d calculations using the $t=90$ snapshot of the 0528a simulation
  ($P_{\rm gas} \sim P_{\rm rad}$).
  The black and blue curves represent calculations that
  exclude and include the effects of Faraday rotation, respectively.
  The dotted curve corresponds to the fitting function given by
  eq. \ref{eq:qfit}.
  \label{f:pinc}}
\end{figure}

Figures \ref{f:mc060} - \ref{f:mc1112} are all plotted for the same
inclination of $79^\circ$, a relatively high value at the edge of the
observed distribution for black hole X-ray binaries \citep[see
e.g.][]{nm05}.  Since polarization increases with inclination,
it will generally be lower in most sources and the effects
discussed here will be more difficult to measure.  The inclination
dependence of $Q$ is shown in Figure \ref{f:pinc}.  The black, blue,
and dashed curves have the same meaning as in the bottom panels of
Figures \ref{f:mc060} - \ref{f:mc1112}.

These results can be compared with previous work on 1d atmospheres
with uniform \citep{sil02,sgs03,gss06} and turbulent \citep{sil07}
magnetic fields.  Our ad hoc relation (\ref{eq:qfit}) is qualitatively
similar to the polarization dependence derived in these works.  They
find depolarization at photon energies with large Faraday rotation
angles and make detailed predictions for the angular and energy
dependence of the polarization.  Their analytic formulas are most
precise in the asymptotic limit of large Faraday rotation angles, but
it is difficult for us to probe the asymptotic dependence due to
limited photon statistics at low energies.  For uniform magnetic
field, they find $p \propto \delta^{-1}$ for $\delta \gg 1$, where
$\delta$ is half the Faraday rotation angle for optical depth of
unity.  This is the same dependence as in (\ref{eq:qfit}) for $\lambda
\gg \lambda_B$.  For isotropic turbulent fields, \citet{sil07}
predicts $p \propto \lambda^{-4}$ for large $\lambda$, a somewhat
stronger dependence than found here.  Although the simulation magnetic
fields do have turbulent fluctuations, there is a mean toroidal field
which may explain our consistency with their uniform field relation.

\section{Full Disk Polarization Model}
\label{diskpol}

The effects discussed in \S\ref{mcpol} indicate that the assumption of
\citet{cha60} polarization for a semi-infinite atmosphere is not
correct at either high or low photon energies.  However, it is
important to keep in mind that our simulations only represent a small
patch of the disk, and (at best) only approximate the emission from a
single, narrow annulus in the accretion flow.  In order to assess the
possibilities for obtaining constraints from actual observations, we
need to compute full disk models.  Given the uncertainties and
complexities involved, our goal is only an approximate model, which is
presented to motivate future calculations and observations.

Following previous work \citep[ and references therein]{hub00}, we
assumed a fully-relativistic thin disk model \citep{ss73,nt73} with
zero torque inner boundary condition at the innermost stable circular
orbit.  We also use the KERRTRANS code \citep{ago97} to compute the
relativistic effects on photon geodesics and parallel transport the
polarization vectors as they travel from the disk surface to the
observer at infinity.  

For the sake of simplicity, we will assume that the local spectra are
color-corrected blackbodies with a color-correction of 1.7, which is a
qualitatively reasonable approximation at the accretion rates of
interest \citep{st95,dav05}.  Of course, the results presented in
\S\ref{tlusty} (and previous work) indicate that this will not be
quantitatively correct in detail.  The main obstacle to a more precise
calculation is that we do not have simulations for the innermost rings
which account for most of the radiation, so we can not robustly model
the effects of dissipation and magnetic support for the whole disk.
Based on the simulations presented in this paper (which, however, may
not generalize to hotter annuli), the overall shift in the spectrum
may be very small, and the numerous approximations (outlined
below) ultimately make this a qualitative enterprise, anyway.

We additionally assume that the local spectra have an angular
distribution that approximately matches the \citet{cha60}
limb-darkening profile, although this only has a small effect on the
output spectrum.  Note that we are also assuming that the local rest
frame polarization is azimuthally symmetric and either parallel or
perpendicular to the disk plane.  We neglect the azimuthal dependence
of polarization position angle discussed in section 5.3 above, due to
the much larger uncertainties inherent in this illustrative full disk
model.

For the energy dependence of the polarization, we use equation
(\ref{eq:qfit}) as an analytic approximation for each individual
annulus.  This accounts for the effects of Faraday rotation, but omits
two other polarization-changing effects that could be important.  If
there is a hot, optically-thin corona above the disk surface, Compton
scattering within the corona could rotate the polarization angle to be
nearly perpendicular to the surface at energies above the thermal peak
\citep{haa93,hb98}.  In the simulation data we study, the gas in the
region above the thermalization photosphere is hotter than the
effective temperature, but the inverse Compton scattering occurring
there is much too weak to explain the coronal emission typically
observed; consequently, our calculations likely underestimate the
importance of this effect, i.e., overestimate the energy at which the
polarization rotation takes place.  Secondly, we do not consider the
scattering of returning radiation \citep{ak00,sk09}, a global general
relativistic effect that also rotates the polarization (and
strengthens it) at energies above the maximum effective temperature
found in the disk.

The last requirement is a specification for $B_{\rm ph}$ as a function
of radius in the disk model.  In the three shearing-box simulations,
$B_{\rm ph}$ correlates well with the equipartition magnetic field
strength $B_{\rm eq}$, the magnetic field strength for which magnetic
pressure equals the total midplane pressure.  Although there is some
scatter in the correlation, we find that $B_{\rm ph} \simeq B_{\rm
  eq}/40$.  Therefore, we estimate $B_{\rm ph}$ by first computing
$B_{\rm eq}$ as a function of radius in our thin disk models and then
use this relation to obtain $B_{\rm ph}$.

This estimate is particularly uncertain, principally because the total
midplane pressure depends on $\alpha$ in the thin disk model, so that
our estimate of $B_{\rm ph}$ inherits this dependence: $B_{\rm ph}
\propto \alpha^{-1/2}$.  We assume $\alpha = 0.01$ in our models, a
value which is $\simeq 1/2$ the typical long-term time-average found
in our stratified shearing box simulations, and roughly an order of
magnitude smaller than both the typical numbers found in the disk body
in global disk simulations and some observationally-based estimates
\citep[e.g.][]{kpl07}.  Using a constant $\alpha$ for this purpose may
create a further problem in that global disk simulations often show a
sizable increase in the time-average of this quantity in the region
just outside the ISCO, where a significant part of the luminosity is
created.  An additional uncertainty is introduced by the fact that
what matters for Faraday rotation is ${\bf B \cdot \hat k}$, and the
relative magnitudes of the different components of ${\bf B}$ may
change systematically with radius.

\begin{figure}[h]
\begin{center}
\hbox{
\psfig{figure=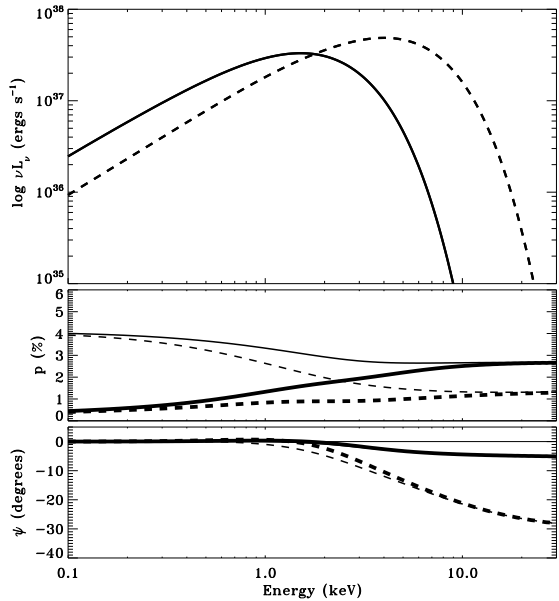,width=0.5\textwidth,angle=0}
}
\end{center}

\caption{Full disk specific intensity (top panel), degree of
  polarization (middle panel) and polarization position angle (bottom
  panel), as viewed by an observer at infinity at an inclination of
  $73^\circ$ to the disk normal.
All spectra are generated from relativistic disk models assuming 
  $L/L_{\rm Edd}=0.1$, $\alpha=0.01$, and $M=10 M_\odot$, which are described
  further in the text.  The solid and dashed curves are computed for $a_*=0$
  and 
$a_*=0.99$, respectively.  In the bottom panels, the thick curves correspond 
to our fiducial model for which eq. \ref{eq:qfit} is used to specify the 
polarization emitted from each annulus and $B_{\rm ph}=B_{\rm eq}/40$ is
assumed.  
The thin curves correspond to models where the polarization is assumed be the
Chandrasekhar value.\label{f:full1}}
\end{figure}

\begin{figure}[h]
\begin{center}
\hbox{
\psfig{figure=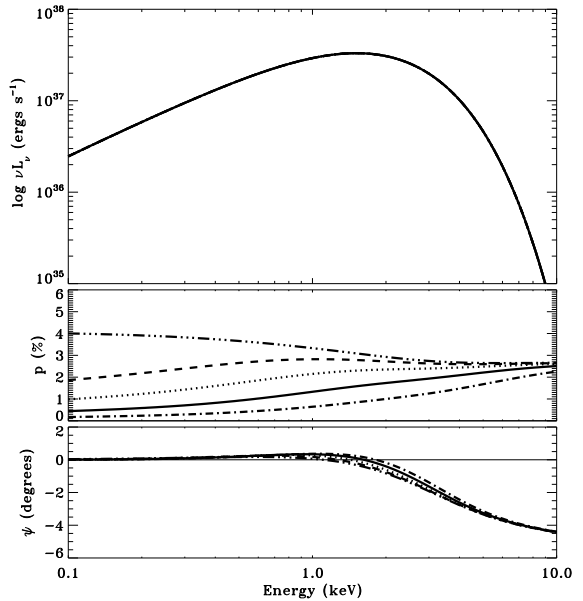,width=0.5\textwidth,angle=0}
}
\end{center}

\caption{Full disk specific intensity (top panel), degree of
  polarization (middle panel) and polarization position angle (bottom
  panel), as viewed by an observer at infinity at an inclination of
  $73^\circ$ to the disk normal.
All spectra are generated from relativistic disk models assuming 
  $L/L_{\rm Edd}=0.1$, $\alpha=0.01$, $M=10 M_\odot$, and $a_*=0$, which are described
  further in the text.  The solid curve corresponds to our fiducial
  model for which $B_{\rm ph}=B_{\rm eq}/40\equiv B_0$.  The other curves correspond
  to different assumption for the magnetic field strength: no magnetic field
  (triple dot-dashed curve), $B_{\rm ph}=B_0/10$ (dashed curve), $B_{\rm ph}=B_0/3$ 
  (dotted curve), and $B_{\rm ph}=3 B_0$ (dot-dashed curve).\label{f:full2}}
\end{figure}

After combining these prescriptions, we compute full disk models, the
results of which are shown in Figures \ref{f:full1} and \ref{f:full2}.
The top panels show the total specific intensity, the middle panels
show the degree of polarization, and the bottom panels show the
polarization angle $\psi$.  Polarization parallel to the plane of the
disk corresponds to $\psi=0^\circ$, while $\psi=90^\circ$ corresponds
to polarization parallel to the projected rotation axis of the disk.

It has been well established by previous work that the polarization is
sensitive to the properties of the spacetime and observer inclination
\citep[see e.g.][]{cs77,sc77,csp80,lnp90,ago97,ak00,dkm04,dov08,lnm09,sk09}.  
Here we briefly review the effect of spin on the polarization results
and refer the reader to previous work for further discussion.  The
solid and dashed curves in Figure \ref{f:full1} are computed for
$a_*=0$ and $a_*=0.99$, respectively.  Increasing the spin parameter has
the well-known effect of shifting the spectral peak to higher energies.

The thin curves show the effects of the spacetime on polarization most
clearly.  These models assume the polarization at the disk surface is
everywhere identical to the \citet{cha60} value.  Because spacetime
near the black hole is strongly curved, photons traveling at an
oblique direction at infinity are in some cases emitted at fluid-frame
directions rather closer to the polar axis.  The parallel transport of
polarization then leads to an effective dilution of the observed
polarization.  This effect is strongest at small radii, where the
highest-energy photons are predominantly created, so the polarization
diminishes somewhat toward higher energy.  There is also a net
rotation of the polarization angle which can be seen in the bottom
panel.  For observer viewing angles as shown, which are less than
$90^\circ$ to the disk rotation axis (defined in a right-handed
sense), the sense of the rotation of the polarization position angle
is clockwise as one moves to higher photon energies \citep{csp80}.
Note, however, that all our calculations omit another relevant
relativistic effect: the scattering of returning radiation.  As shown
by \citet{ak00} and in greater detail by \citet{sk09}, this effect can
lead both to a {\it rise} in polarization at high energies and a
nearly $90^\circ$ rotation of its direction.

The thick curves in the bottom two panels of Figure \ref{f:full1} show
the results from disk models for which equation (\ref{eq:qfit})
specifies the polarization at the disk surface.  The low energy
polarization is significantly reduced by Faraday depolarization, and
is negligible at or below 0.1 keV.  The degree of polarization
increases with photon energy, but even near the peak, the polarization
is reduced significantly from the non-magnetized, Thomson scattering
models.

From Figure \ref{f:full1} it is clear that Faraday rotation has a
significant effect for the magnetic field strengths adopted in our
model.  However, it may be the case that our assumed
$B_{\rm ph} \simeq B_{\rm eq}/40 \equiv B_0$
does not generally hold or that our estimate of
$B_{\rm eq}$ from the thin disk model may either be higher or lower
than in real flows.  Therefore, we explore the sensitivity to this
assumption in Figure \ref{f:full2} by plotting several spectra with
identical spin, mass, accretion rate, and inclination, but with
different assumptions for magnetic field strength.

At the high energy end, all the models asymptote to a similar polarization
set by the Chandrasekhar value for this inclination, but with a slight
reduction due to relativistic effects.  The effects of varying the
magnetic field are seen at energies near the spectral peak (in $\nu
I_\nu$) and at low energies.  Quite generally, the level of
polarization decreases at all energies as we strengthen the magnetic
field.  The wavelength dependence is also sensitive to field strength.
The model with $B_{\rm ph}=0$ (triple-dot-dashed curve) shows a
monotonic decline in polarization as photon energy increases.  In the
models with lower but non-zero magnetic field ($B_{\rm ph}=B_0/10$,
$B_0/3$), the degree of polarization generally increase as photon
energy increases and reaches a maximum or begins to level off at or
just below the spectral peak.  For even higher fields ($B_{\rm
  ph}=B_0$, $3 B_0$), the polarization continues rising and only
levels off out in the Wien tail.  

We can compare our results with those of \citet{gss06}, who also
examined the wavelength dependence of polarization in accretion disks subject
to Faraday rotation.  They computed the polarization from accretion
disks with purely vertical magnetic field using the analytic models of
\citep{sil02}, but neglecting the effects of relativistic transfer.
They reported the long wavelength asymptotic dependence of the
polarization for various assumptions about radial dependence of the
magnetic field.  We find an approximate asymptotic dependence of $p
\propto \lambda^{-s}$ with $s \sim 0.3$.  Long wavelength
photons are predominantly emitted at larger radii where relativistic
effects are negligible and $P_{\rm gas} > P_{\rm rad}$, so $B_{\rm ph}
\propto B_{\rm eq} \propto r^{-5/4}$ \citep{ss73}.  For this radial
dependence, \citet{gss06} find $s = 1/3$ in reasonable agreement
with our result.  The agreement is not surprising since the depolarization
parameter (their $\delta$) is not very sensitive to the field geometry,
and the asymptotic dependence found by \citet{sil02} is very similar to
that of equation (\ref{eq:qfit}).

\section{Discussion}
\label{discus}

\subsection{Implications for Spectral Hardening and Continuum Spin Estimation}

One of the methods currently being used to try and measure the spins of
black holes is fitting the thermal X-ray continuum data to accretion disk
models, either directly to the models themselves \citep{ddb06,mid06}
or to color-corrected blackbody disks using color correction factors
measured from the models \citep{sha06,mcc06,liu08}.  All other things
being equal, a (prograde) disk around a more rapidly spinning black hole
should produce a harder spectrum than around a non-rotating back hole.
In order for this method to be viable, the intrinsic hardness of the locally
emitted spectrum, which deviates significantly from a blackbody at the same
effective temperature because of electron scattering, must be accurately
determined from theory.

All spectral models to date neglect the effects we have been exploring
in this paper.  In particular, magnetic vertical support hardens the spectrum
and density inhomogeneities soften the spectrum.  In the three simulation
snapshots we examined, these two effects nearly canceled.  Remarkably,
this cancellation occurred even though the hardening/softening due to
both effects ranged from $\sim 3-12$\% across the different snapshots.
This is suggestive of a more general result, which could arise because
the magnetic fields providing the support against gravity also generate
the density inhomogeneities.  Nevertheless, we caution that it may be still be
the case that these cancellations were fortuitous, and that
in other regimes one or the other of these two effects will dominate.

While we have achieved some success in quantifying the uncertainties
in the color correction factors of the locally emitted spectrum, it
should be borne in mind that continuum spectral fitting methods to
measure black hole spin are also subject to uncertainty in the radial
emissivity profile of the disk.  See \cite{bhk08}, \cite{sha08}, and
\cite{nkh09} for recent discussions of this issue.

\subsection{Polarization Predictions and Potential Magnetic Field Measurements}

The polarization dependence on magnetic field strength displayed in
Figure \ref{f:full2} demonstrates that polarization can, in principle,
be used to constrain the magnetic field strength in real accretion
flows.

However, it is important to keep in mind that we have neglected
several effects.  First, we do not include the effect of absorption,
which will also tend to decrease the polarization at the lowest
energies, as discussed in \S\ref{mcpol}.  The free-free absorption
opacity included in the MC calculations produces a drop in
polarization which occurs roughly half a decade lower in energy than
the Faraday depolarization effect.  This is probably an underestimate
of the effect, since bound-free and bound-bound opacity can be
significantly greater than free-free opacity for energies in the 0.1-2
keV range.  The bound-free opacity is clearly significant in the 0326c
and 0528a equivalent 1d spectra (Figs. \ref{f:s060} and \ref{f:s090},
respectively).  However, for the hotter annuli which dominate the
output near the spectral peak, the ionization state is sufficiently
high that bound-free opacity has at most a modest effect (see e.g.
Fig. \ref{f:s1112}).

Faraday rotation and absorption opacity can occasionally interact in
subtle ways to {\it increase} the emergent
polarization, depending on the vertical gradient of the thermal source
function in the atmosphere.  The physics behind this is
discussed extensively by \citet{abi98}, who demonstrated that this
can happen near the Balmer edge in disk atmosphere models appropriate
for quasars.  It is possible that such effects may also occur near
bound-free absorption edges in X-ray binary disks.

Although astronomical X-ray spectropolarimetry has hitherto been
limited to observations of the Crab Nebula \citep{wei76},
there is now a realistic possibility of observing these polarization
effects in Galactic black holes.  The {\it Gravity and Extreme
  Magnetism Small Explorer} mission \citep{jah07,swa08} is now
scheduled for launch in 2014 and promises to have both the sensitivity
and energy resolution in the 2--10 keV band to detect these effects in
a number of objects.   Similar technologies are also being
considered for an X-ray polarimetry instrument on board the future
{\it International X-ray Observatory}.

\subsection{Generality of these predictions}

If our assumed scaling of $B_{\rm ph}$ with $B_{\rm
  eq}$ generally holds, Faraday rotation is likely to be important in
most luminous accretion regimes.  The innermost regions of near
Eddington accretion disks are radiation dominated for $\sim 10 \msun$
black holes, and are even more radiation dominated for supermassive
black holes.  Using the radiation dominated relations of \citet{ss73}
we can compute the characteristic rotation angle for optical depth
unity near the spectral peak using equation (\ref{eq:farad})
\be
\chi_{\rm F,p} \simeq \frac{3 \lambda_{\rm p}^2}{16\pi^2 e} {\bf B_{\rm ph} 
\cdot {\hat k}},
\label{eq:chi2}
\ee
where $\lambda_{\rm p} \simeq c h/(4 k_{\rm B} T_{\rm eff})$.  The effective
temperature $T_{\rm eff}$ is computed from the flux
\be
T_{\rm eff} = \left(\frac{3 c^3 \dot m R_{\rm R}}
                  {2R_g \kappa_{\rm es} \eta \sigma r^3}\right)^{1/4} ,
\ee
where $R_g=GM/c^2$ is the gravitational radius, $\kappa_{\rm es}$ is
the electron scattering opacity, $\sigma$ the Stefan-Boltzmann
constant, $r$ is the radius in gravitational radii, $\eta$ is the
radiative efficiency, $\dot m$ is the accretion rate normalized to the
Eddington rate $\dot{M}_{\rm Edd}=4\pi R_g c/(\kappa_{\rm es} \eta)$,
and $R_R$ is a function encapsulating both relativistic corrections
and the ISCO torque boundary condition (notation following
\citealt{kro99}).  Defining $B_{\rm eq} = (8\pi P_{\rm rad})^{1/2}$
and using the results of \citet{ss73}, we have
\be
B_{\rm ph} = \frac{B_{\rm eq}}{40} = 
\left(\frac{8\pi}{\alpha\kappa_{\rm es}R_g}\right)^{1/2}
\frac{c}{40 r^{3/4}}\left(\frac{R_{\rm z}R_{\rm T}}{R_R}\right)^{1/2},
\label{eq:bph}
\ee
where $R_{\rm T}$, like $R_{\rm R}$, summarizes both relativistic and
boundary condition corrections to the stress profile, and $R_{\rm z}$
is the relativistic correction to the vertical gravity.  Combining
equations (\ref{eq:chi2})-(\ref{eq:bph}) with the definition of
$\lambda_{\rm p}$ we arrive at
\be
\chi_{\rm F,p} \sim 1 \left(\frac{\eta}{\dot m}\right)^{1/2}
                       \frac{r_{10}^{3/4}}{\alpha_{0.01}^{1/2}q_4^2 Q_{40}}
                      \frac{(R_{\rm z} R_{\rm T})^{1/2}}{R_{\rm R}}
                       {\bf \hat b \cdot \hat k} ,
\ee
where ${\bf \hat b}$ is a unit vector in the direction of the mean
field, $q = hc/(k_B T_{\rm eff} \lambda_{\rm p})$, and $Q = B_{\rm
  eq}/B_{\rm ph}$.

Remarkably, the characteristic Faraday rotation angle is {\it
  independent} of black hole mass wherever the disk is dominated by
radiation pressure.  Moreover, at least to the accuracy of this
estimate, it is always $\sim O(1)$ for wavelengths near the thermal
peak for the radii where most of the light is produced when $\dot m
\sim \eta$.  Although the scaling we have derived should be
robust, the actual magnitude of the effect is subject to significant
uncertainty.  As we have already discussed, the magnitude of $\alpha$
could likely be somewhat larger than the fiducial value we have chosen
for it.  Color corrections could alter $q$.  Although simulations at a
variety of radii have all pointed toward $Q \simeq 40$, we cannot say
for certain whether this ratio applies in the innermost disk.  The
particular radius dominating the light output depends on black hole
spin and the ISCO boundary condition.  Because $r \simeq 10$ is
appropriate for non-rotating black holes with little torque at the
ISCO, spin and extra stress in the inner disk would diminish the
characteristic rotation angle, both by decreasing the characteristic
radius of emission and by increasing $R_{\rm R}$ at small radii
(typically $R_{\rm T} \simeq R_{\rm R}$, while $R_{\rm z} \gtrsim 1$).

We therefore predict that Faraday rotation {\it could} substantially
affect the emergent polarization both for high-accretion rate Galactic
black holes (in the thermal or steep power-law states) and for many
AGN (as has been previously pointed out by, e.g., \citealt{bla90}).
Prospects for using this effect to constrain disk field properties are
better in the stellar-mass case than in the supermassive case,
however, because in AGN it appears that the observed polarization does
not arise from the accretion disk itself (e.g. \citealt{ant88}), but
from scattering by material further out.  Accretion disks in
cataclysmic variables also emit the bulk of their light in the optical
and ultraviolet, but these disks are not scattering dominated.  The
theoretically predicted continuum polarization in the outburst phase
is very small \citep{che88}, and has in fact defied observational
attempts to separate it out from interstellar polarization
\citep{nay96,mof01}.  X-ray polarimetry of X-ray binaries might
therefore be an optimal way of constraining accretion disk magnetic
fields.

\section{Conclusions}
\label{conc}

Using snapshots of local shearing box simulations of accretion disks
with a broad range of radiation to gas pressure ratios, we have
calculated how a realistic vertical structure established by MRI
turbulence affects the emergent photon spectrum and polarization.
Most of the dissipation within the turbulence occurs at high effective
optical depth in all the physical regimes that have been simulated
thus far.  As a result, the spectra are very insensitive to the
details of that dissipation profile, and are extremely close to
standard model spectra that assume that the vertical profile of
dissipation per unit mass is constant.  On the other hand, standard
models usually neglect the fact that the photospheric regions are
supported against gravity by magnetic forces.  These forces reduce the
horizontally-averaged densities at the thermalization photosphere,
resulting in harder 1d model spectra compared to standard models.

In addition to lifting the atmosphere, those same magnetic forces also
produce complex three-dimensional density inhomogeneities in the
photospheric regions.  Through a comparison between 3d Monte Carlo
calculations and 1d calculations that were designed to have the same
emergent flux, we demonstrated that these inhomogeneities act to
soften the spectrum.  Somewhat surprisingly, this softening largely
cancels the hardening due to vertical magnetic support in all three
snapshots, even though magnitude of each effect differs between the
three simulations. Therefore, it may be that the color corrections
derived from, e.g. BHSPEC models \citep{dh06}, will be approximately
correct.  However, we cannot rule out the possibility that this
cancellation was fortuitous in the three snapshots we considered here.
Given the magnitude of the effects we have found, we caution observers
that color correction factors derived from atmosphere models that
neglect magnetic support and 3d inhomogeneities may be incorrect by
approximately ten percent in either direction.

Because the density inhomogeneities produce a complex, structured
photosphere, the degree of polarization of the emerging radiation
field is reduced compared to a plane-parallel atmosphere, but only
slightly.  Faraday rotation by the strong magnetic fields in the
atmosphere produces a much stronger effect, producing significant
reduction in polarization for photon energies near or below the peak
in the local spectrum.  This fortunate coincidence that Faraday
rotation is strong, but not too strong, means that future X-ray
polarimetry measurements of the polarization of the thermal component
of black hole X-ray binaries could be used as a diagnostic of magnetic
fields in disks.  More extensive theoretical calculations will be
necessary, however, before this can be done quantitatively.

\acknowledgements{
We thank J. Schnittman for useful discussions.
This work was supported in part by NSF grant AST-0707624. SD is supported by
NASA grant number PF6-70045, awarded by the Chandra X-ray Center,
which is operated by the Smithsonian Astrophysical Observatory for
NASA under contract NAS8-03060.
}

\appendix

\section{Implementation of Polarized Compton Scattering in the Monte Carlo
Method}

\cite{blp82} provide an incisive summary and first-principles
derivation of the cross sections for both polarized and unpolarized
Compton scattering.
Polarized radiation can be completely described by a Stokes vector
${\bf I} \equiv (I,Q,U,V)^{\rm T}$, where $I$ is the total intensity,
$Q$ and $U$ are intensities of linearly polarized radiation, and $V$ is the
intensity of circular polarization \citep{cha60}. We only consider
polarization arising from electron scattering, so $V=0$ and we will
drop it in what follows.  Since we only consider scattering of
individual photon packets, it will be convenient to work in terms of
normalized Stokes parameters $q \equiv Q/I$ and $u \equiv U/I$.  Note
that these quantities correspond to $\xi_3$ and $\xi_1$, respectively,
in the notation of \cite{blp82}.

We consider the general problem of modeling electron scattering in the
comoving frame of the fluid (as opposed to the electron rest frame).
We define photon and electron momentum four vectors
\be
K_\mu &=& \omega m_{\rm e} c (1,{\bf k})  \nonumber \\
P_\mu &=& \gamma m_{\rm e} c (1,{\bbeta}),
\ee
where $\omega \equiv h \nu/(m_{\rm e} c^2)$ and all other variables
have their usual meanings.
It is convenient to work with relativistic invariants $x\equiv 2 P^\mu
K_\mu/(m_{\rm e} c)^2$ and $x' \equiv 2 P^\mu K_\mu'/(m_{\rm e} c)^2$,
where primes denote the scattered quantities.
Conservation of momentum then requires
\be
x'=x - 2 \omega \omega' (1- \cos \theta),
\ee
where $\cos \theta \equiv {\bf k} \cdot {\bf k}'$ is fluid frame
scattering angle.

The differential scattering cross section can be summed over final
photon polarizations to yield
\be
\frac{d \sigma}{d \Omega'}=\frac{3 \sigmat}{16 \pi}\left(\frac{x'}{x}\right)^2
\left[\left(\frac{x'}{x}+\frac{x}{x'}\right)-4 \delta (\delta+1)(1-q_{\rm in}) \right],
\label{eq:xsec}
\ee
where $\delta \equiv 1/x-1/x'$ and $q_{\rm in}$ is the Stokes parameter evaluated 
in the ``internal'' basis defined by
the scattering plane.  In general this differs from the ``external''
basis defined relative to the fluid frame.  They are related to each
other through rotation matrices $L(\chi)$, which transform ${\bf I}$ to
the internal basis before scattering, and $L(-\chi')$, which transforms
${\bf I'}$ back to the external basis after scattering.  The matrices
are given by \citep{cha60}
\be
L(\chi)=\left(\begin{array}{ccc}
1 \;& 0 \;& 0 \\
0 \;& \;\;\cos{2\chi} \;& \sin{2\chi} \\
0 \;& -\sin{2\chi} \;  & \cos{2\chi}  \end{array}\right) \label{eq:rot}
\ee
Using these rotation matrices, one finds $q_{\rm in}=q\cos 2\chi - u
\sin 2\chi$

\cite{nap93} provide a thorough discussion of polarization bases and derive
expressions for the angles $\chi$ and $\chi'$ in their appendices.
Here, we just quote their results
\be
\cos \chi & = & \frac{\gamma(k_z'-\cos \theta k_z)-\beta \gamma
\left[(1-\cos \theta) \hat{\beta}_z+ {\bf k \cdot} \hat{\bbeta} (k_z'-k_z)\right]}
{\sqrt{1-k_z^2} \sqrt{(\delta-1)/\delta} (1-\cos \theta)}, \nonumber \\
\sin \chi & = & \frac{\gamma(1-\bbeta {\bf \cdot  k})(k_y k_x'-k_x k_y')
-\beta \gamma (1-\cos \theta)(\hat{\beta}_x k_y-\hat{\beta}_y k_x)}
{\sqrt{1-k_z^2} \sqrt{(\delta-1)/\delta} (1-\cos \theta)}, \label{eq:chi} \\
\cos \chi' & = & \frac{\gamma(k_z-\cos \theta k_z')-\beta \gamma
\left[(1-\cos \theta) \hat{\beta}_z+ {\bf k' \cdot} \hat{\bbeta} (k_z-k_z')\right]}
{\sqrt{1-k_z'^2} \sqrt{(\delta-1)/\delta} (1-\cos \theta)}, \nonumber \\
\sin \chi' & = & \frac{\gamma(1-\bbeta {\bf \cdot k'})(k_y' k_x-k_x' k_y)
-\beta \gamma (1-\cos \theta)(\hat{\beta}_x k_y'-\hat{\beta}_y k_x')}
{\sqrt{1-k_z'^2} \sqrt{(\delta-1)/\delta} (1-\cos \theta)}. \label{eq:chip}
\ee
Here, $\hat{\beta} \equiv \bbeta/\beta$ is a unit vector in the electron 
momentum direction.

The effects of scattering on the Stokes vector are accounted for by the
matrix $R=L(\chi')SL(-\chi)$ where $F$ is the transformation induced
by Compton scattering in the internal basis \citep[see
e.g.][]{nap93,blp82}:
\be
\bf{S}=\left(\begin{array}{ccc}
S_a+S_b & S_c & 0 \\
S_c & S_a & 0 \\
0 & 0  & S_d  \end{array}\right), \label{eq:smat}
\ee
where $S_a=4 \delta (\delta+1)+2$, $S_b=(x'/x+x/x')-2$, $S_c=S_a-2$,
and $S_d=4 \delta+2$.  In the $\beta, \beta' \rightarrow 0$ limit, $x'
\rightarrow 2\omega'$, $x \rightarrow2\omega$, and $\delta \rightarrow
(\cos \theta - 1)/2$ so that $S_a \rightarrow \cos^2 \theta - 1$, $S_b
\rightarrow 0$, $S_c\rightarrow \cos^2 \theta + 1$, $S_d \rightarrow 2
\cos \theta$.  Therefore, $\bf{S}$ corresponds to the Rayleigh matrix
for Thomson scattering in the appropriate limit. Carrying out the
matrix multiplication, we find
\be
\bf{R}=\left(\begin{array}{ccc}
S_a+S_b & S_c \cos 2\chi & -S_c \sin \chi \\
S_c \cos 2\chi' & S_a \cos 2\chi\cos 2\chi'+ S_d \sin 2\chi \sin 2 \chi' & 
-S_a \sin 2\chi\cos 2\chi'+ S_d \cos 2\chi \sin 2 \chi'\\
-S_c \sin 2\chi' & -S_a \cos 2\chi\sin 2\chi'+ S_d \sin 2\chi \cos 2 \chi'  &
S_a \sin 2\chi\sin 2\chi'+ S_d \cos 2\chi \cos 2 \chi'
\end{array}\right) \label{eq:rmat}.
\ee

\subsection{Monte Carlo Implementation}

\cite{pss83} (hereafter PSS83) describe and evaluate Monte Carlo
methods for Compton scattering of unpolarized radiation.  Using the
above relations, it is straightforward to generalize the methods
presented in PSS83 to include polarization.  For the sake of brevity,
we focus only on the additional complexity offered by the polarization
physics and refer the reader to PSS83 for further details.

The angle integrated cross section $\sigma(x)$ is independent of
polarization, and equivalent to eq. (2.10) of PSS83.  Therefore, the
mean free path and selection of the electron momentum can be evaluated
using the exact same method as presented in sections 9.4 and 9.5 of
PSS83.  If the photon momentum is accepted, the scattered photon
direction ${\bf k'}$ is then drawn randomly, and $\cos \theta$ and
$x'$ are evaluated as before.  However, the scattering probability
(eq. 9.8 of PSS83) now depends on the Stokes parameters through eq.
(\ref{eq:xsec}).  The quantity $X$ (defined in \S 2.1.2 of PSS83) is
replaced by the quantity in square brackets in eq. (\ref{eq:xsec}).
Therefore, we must evaluate $q_{\rm in}$, which requires calculation of
$\chi$ via eqs. (\ref{eq:chi}).

If accepted, the photon direction and energy are updated as in PSS83.
Additionally, the normalized Stokes parameters after scattering must
be calculated.  Using $\bf{I'} = \bf{R \cdot I}$ and eq. \ref{eq:rmat}
we find
\be
q' & = & \frac{R_{21} + R_{22} q + R_{23} u}
{R_{11} + R_{12} q +R_{13} u}, \nonumber \\
u' & = & \frac{R_{31} + R_{32} q + R_{33} u}
{R_{11} + R_{12} q +R_{13} u}.
\ee

\end{document}